\documentclass[12pt]{article}
\usepackage{a4wide}
\usepackage{graphicx,amsmath,amssymb}

\newcommand{\be}{\begin{equation}}
\newcommand{\ee}{\end{equation}}
\newcommand{\bea}{\begin{eqnarray}}
\newcommand{\eea}{\end{eqnarray}}
\newcommand{\eref}[1]{(\ref{#1})}
\newcommand{\rmi}{{\rm i}}
\newcommand{\rme}{{\rm e}}

\newcommand{\etas}{\eta_0}

\begin{document}

\title{The Robustness of $n_s \lesssim 0.95$ in Racetrack Inflation}

\date{December 4, 2007}

\author{Ph.~Brax$^1$\footnote{brax@spht.saclay.cea.fr}, 
S.~C.~Davis$^2$\footnote{sdavis@lorentz.leidenuniv.nl}, 
and M.~Postma$^{3,4}$\footnote{postma@mail.desy.de}
\\ \\ \em
${}^1$ Service de Physique Th\'eorique, CEA/DSM/SPhT, \\ \em
Unit\'e de recherche associ\'ee au CNRS, \\ \em 
CEA--Saclay 91191 Gif/Yvette cedex, France
\\ \\ \em
${}^2$ Laboratoire de Physique Theorique d'Orsay, B\^atiment 210, \\ \em 
Universit\'e Paris-Sud 11, 91405 Orsay Cedex, France
\\ \\ \em
${}^3$ DESY, Notkestra\ss e 85, 22607 Hamburg, Germany
\\ \\ \em
${}^4$ Nikhef, Kruislaan 409, 1098 SJ Amsterdam, The Netherlands
}

\maketitle

\begin{abstract}
A spectral index $n_s \lesssim 0.95$ appears to be a generic prediction of
racetrack inflation models.  Reducing a general racetrack model to a
single-field inflation model with a simple potential,  we obtain an analytic
expression for the spectral index, which explains this result.  By considering
the limits of validity of the derivation, possible ways to achieve higher
values of the spectral index are described, although these require further
fine-tuning of the potential. 
\end{abstract}

\hfill {DESY 07-210, LPT-ORSAY-07-124}

\section{Introduction}

Racetrack inflation~\cite{race1} is an explicit realization of modular
inflation within the context of KKLT volume stabilisation~\cite{kklt}.  It
employs a superpotential of the double-exponential racetrack form. The
inflaton is then the imaginary part of a geometric modulus.  Inflation begins
near a saddle in the moduli potential, and ends when the inflaton fast rolls
towards a local minimum of the potential.  A big success of this class of
models is the seemingly very robust prediction $n_s \lesssim 0.95$ for the
spectral index, which is very close to the latest WMAP
results~\cite{WMAP3}. In this letter we explain the origin of this upper bound
on the spectral index, which is a generic feature of double- as well as
many-exponential superpotentials.  In addition, we will show that to a very
good approximation the spectral index only depends on one parameter, namely
the value of the slow roll parameter $\eta$ at the saddle point.

\section{Racetrack inflation}

Modular inflation occurs in the KKLT scenario at sufficiently flat saddle
points of the potential. There the real part of the volume modulus
$T$ is stabilised, and the tachyonic, imaginary part plays the role of the
inflaton. The KKLT~\cite{kklt} racetrack potential comes from a supergravity
model with
\be K= -3 \ln (T+ \bar T) \, , \qquad
 W= W_0 + A \rme^{-a T} + B \rme^{-b T} \, , 
\label{racetrack1}
\ee 
together with an non-supersymmetric lifting term $V_{\rm lift} = 2^\alpha
E/(T+\bar T)^\alpha$. The constants $a,b$ depend on the specifics of the
non-perturbative physics, which can come from gaugino condensation or
Euclidean instantons. The lifting term is adjusted to obtain a
Minkowski vacuum with a vanishing cosmological constant. Notice that
the lifting term breaks supersymmetry explicitly. It can be realised
in a string context by putting an anti-brane in the bulk
($\alpha = 2$) or at the tip of a warped throat ($\alpha =3$).
Defining $T= X+ \rmi Y$, the resulting potential is
\bea
&&V = \frac{E}{X^\alpha} + \frac{1}{6 X^2}
\biggl\{ A^2 a[aX + 3]\rme^{-2a X} +B^2  b[bX + 3] \rme^{-2b X} +3
W_0 A a \rme^{-a X} \cos(aY) \nonumber \\  && \hspace{.3in} {} +3
W_0 B b \rme^{-b X} \cos(b\,Y) + A B [2abX +3(a+b) ]
\rme^{-(a+b)X} \cos([a-b] Y) \; \biggr\}  \, . 
\label{Vrace} 
\eea
It should be noted that the kinetic terms ${\mathcal L}_{\rm
kin}=(3/4X^2)[(\partial X)^2 +(\partial Y)^2]$ are non-canonical. The
axion field $Y$ is the inflaton~\cite{race1}.  The
potential is periodic in $Y$, giving rise to saddle points between
degenerate local minima.  If its initial value is close enough to the saddle
point, the rolling of the $T$ modulus along the
unstable $Y$ direction produces inflation (see figure 1 in \cite{race1} for
the shape of the potential).

In \cite{race2} an improved racetrack inflation model was proposed with two
geometric moduli:
\be
K= -2\ln\Big[ \frac{  (T_2-\bar T_2)^{3/2} - (T_1-\bar T_1)^{3/2} }{36} \Big]
\, , \qquad
W= W_0 + A \rme^{-a T_1} + B \rme^{-b T_2} \, .
\label{racetrack2}
\ee
In this case the inflaton is a linear combination of the two axions
(the orthogonal combination is a flat direction), and the inflaton
potential has a structure similar to \eref{Vrace}.

Alternatively, racetrack inflation can be obtained without the need
for non-supersymmetric terms by using $D$-term uplifting~\cite{ana}.
Gauge invariance requires additional meson fields $\chi$.  The model is
\be K= -3 \ln (T+
\bar T) +\vert \chi\vert^2 \, , \qquad W= W_0 + A \chi^{-ra}
\rme^{-a T} + B \chi^{-rb} \rme^{-b T} \, .
\label{racetrackD}
\ee 
The $D$-term potential is similar in form to $V_{\rm lift}$, and serves to
uplift the minimum to a Minkowski vacuum.
Although the form of the potential is more complicated~\cite{raceD}, its
structure of minima and maxima is similar to \eref{Vrace}. In particular,
there are still saddle points whose unstable direction is almost
coincident with the $Y$ direction. Like the field $X$, the meson field $\chi$
remains roughly constant during inflation.  This was found to be the case for
all models and parameters studied in~\cite{raceD}.

All these different realizations of racetrack inflation are
effectively single field inflation models with the axion field as the
inflaton.  Although other fields are present --- the $X_i$ and the
meson fields --- they are fixed during the period of inflation when
WMAP scales leave the horizon.  One can therefore approximate the effective
inflaton potential as
\begin{equation}
V(Y)=V_0 + \sum_i A_i \cos a_i Y \, ,
\label{Vcos}
\end{equation}
where the overall phase is chosen such that the saddle/maximum is at
$Y=0$. For the double-exponential superpotentials discussed above, $i=3$. More
generally, the superpotential can have more than two exponentials, and
$i > 3$\footnote{In model~\eref{racetrack2} there is more than one
axion if there are more than two moduli fields.  In this case the
potential only reduces to the form~\eref{Vcos} in the limit of single
field inflation.}. Inflation can occur when the saddle point is such that
 $\etas= V''(Y)/V(Y)\vert_{Y=0}$ is much smaller than one,
guaranteeing that the slow roll conditions are satisfied close enough
to the saddle point. In the following, we will use a further approximation of
this effective potential to show that, quite generically in racetrack
inflation, the spectral index is determined by $\etas$ only, and is bounded
from above.

\section{Inflationary dynamics}

For all the racetrack models mentioned in the previous section, the
physics which determines the cosmic microwave background parameters occurs
close to the saddle point at $Y=0$, where the inflaton is the only field that
evolves significantly.  Taylor expanding the potential at the saddle gives
\be V =
V_{\rm sad}\left( 1 + \etas \frac{y^2}{2} + C \frac{y^4}{4}+\cdots \right)
\, , \label{Vapp} 
\ee
where $\etas$ is the value of $\eta$ at the saddle point, and $y$ is
the canonically normalised inflaton field, which e.g.\ for the racetrack
models (\ref{racetrack1}, \ref{racetrackD}) is $y = \sqrt{3/2} \, Y/X$.
Using this approximate potential we can calculate the number of
e-folds before the end of inflation $N=-\ln a$, as a function of $y$ in the
slow roll approximation:
\be N(y) = \int^y_{y_{\rm end}} \frac{V
\,dy}{V'} \approx \left[ \frac{1}{2\etas}\log\frac{y^2}{y^2 +
\etas/C}\right]^y_{y_{\rm end}} \, . \label{Napp} 
\ee
Inflation ends when the slow roll parameter 
$\epsilon = (1/2)(V'/V)^2 \approx 1$, which for the above potential
happens at $y_{\rm end} \sim C^{1/3}$. Inverting the above equation to get
$y(N)$, and substituting it into the expression for the slow roll parameter
$\eta = V''/V$, we obtain
\be
\eta  \approx  \etas -3 \etas\left(1-\rme^{-2\etas N}
  -\frac{\etas}{C y_{\rm end}^2} \rme^{-2\etas N}\right)^{-1} \, .
\ee
In the racetrack models $|\etas| \ll 1$ is tuned small, whereas the
coefficient $C$ is not.  Hence we can expand the above
expression in $ |\etas/(C y_{\rm end}^2)| \ll 1$. To lowest order
the spectral index $n_s \approx 1+2\eta$ is then
\be
n_s = 1+ 2 \etas -\frac{6 \etas}{1-\rme^{-2N \etas}}
\label{nsapprox}
\ee
evaluated $N=N_* \approx 55$ e-folds before the end of inflation. For the
parameters used in the original racetrack model~\cite{race1} 
$\etas \approx -0.0061$, $C \approx 293$ and $y_{\rm end} \approx 0.12$, while
for the $D$-term uplifted racetrack~\cite{raceD} $\etas \approx -0.0095$, 
$C \approx 999$ and $y_{\rm end}\approx 0.10$. Hence our expansion in 
$\etas/(C y_{\rm end}^2)$ is justified.  In fact, it is a particularly good
approximation since the error in the spectral index from neglecting higher
order terms is only of the order $10^{-4}$. As a result $n_s$ is
practically a function of $\etas$ (and $N_*$) only.

\begin{figure}
\centerline{\includegraphics[width=10cm]{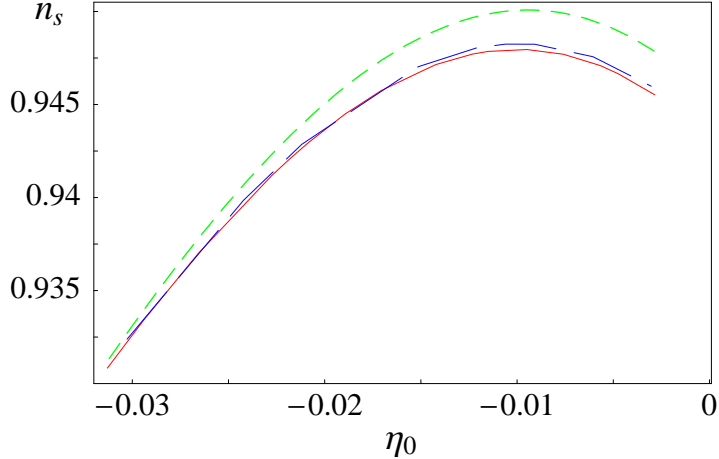}}
\caption{Plot of $n_s(\eta_0)$ for Taylor expanded, approximate cosine, and
  full racetrack potentials (curves from top to bottom).}
\label{F:ns}
\end{figure}

\section{Discussion}

Figure~\ref{F:ns} shows $n_s(\etas)$ with $N=55$ for $D$-term uplifted
racetrack inflation~\eref{racetrackD}, for the effective potential~\eref{Vcos},
and for our analytic result~\eref{nsapprox}.  The results for the
cosine potential are nearly identical to the full racetrack model,
confirming that the dynamics of the ``spectator fields'' ($X_i$ and
$\chi$, which only evolve significantly towards the end of inflation) have a
negligible effect on the inflationary predictions.  Freezing the
spectator fields all the different models proposed
(\ref{racetrack1}, \ref{racetrack2}, \ref{racetrackD}) reduce to the same
effective model, so it is not surprising they give the same inflationary
predictions. Hence the predictions of racetrack inflation are very robust.

The approximation \eref{Vapp} of the full racetrack potential is simple yet
very effective.  It gives $n_s(\eta_0)$ in very good agreement with the
results of the full racetrack potential, the error in $n_s$ is of the order
$10^{-3}$.  Accordingly it correctly predicts an upper bound on the spectral
index $n_s \lesssim 0.95$ for generic cases, i.e.\ with the coefficient $C$
not fine-tuned.  It further explains the puzzling result of the racetrack
models that the spectral index is a function of $\etas$ only; the dependence
on $C$ and $y_{\rm end}$ drops out in the $C y_{\rm end}^2 \gg |\etas|$ limit.
The reason for the small disagreement between the analytic results and the
full potential is that although \eref{Vapp} is a very good approximation when
observable scales leave the horizon, it becomes worse towards the end of
inflation when $y$ is larger.  To improve upon our results we should take
higher terms into account when calculating $N(y)$, which only play a role in
the integration region near $y_{\rm end}$.  We reemphasise that higher order
corrections in $1/C$ only affect the spectral index at the level of $10^{-4}$,
and are not the cause of the small mismatch between the analytic approximation
and the racetrack model.

Our analytical model also suggests ways to get around the upper bound $n_s
\lesssim 0.95$. These require extra tuning, in addition to that needed to get
$|\etas| \ll 1$. One possibility is if inflation is multi-field, in which case
our approximation~\eref{Vcos} breaks down. This is probably most easily
achieved in a set-up analogous to~\eref{racetrack2}, with additional moduli
fields, and so more than one axion field. However, the parameters need to be
tuned to arrange that the saddle point has more than one unstable directions
with similar curvature if multiple  axion fields are to act as inflatons.

\begin{figure}
\centerline{\includegraphics[width=10cm]{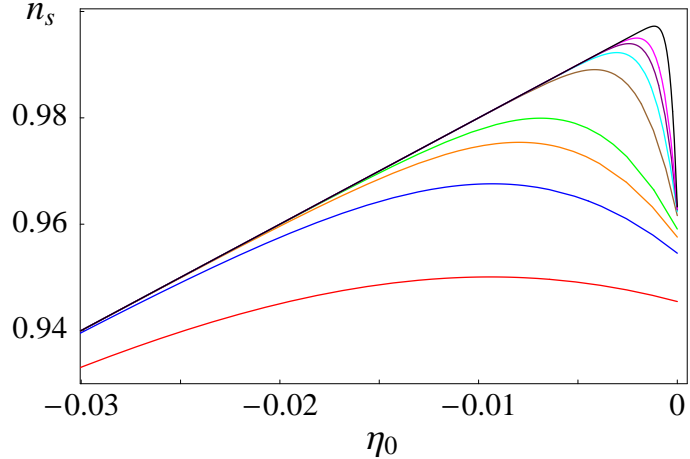}}
\caption{Plot of $n_s(\eta_0)$ given in \eref{nsp} for
$2p = 4,6,8,10,20,30,40,50,100$ (increasing $p$ corresponds to larger
spectral index).}
\label{F:p}
\end{figure}

An alternative way to avoid the upper bound is by fine-tuning the coefficient
of the quartic term in the expansion~\eref{Vapp}. Setting $C\approx 0$ would
lead to a model where the $y^6$ term is dominant. More generally we can
consider a model with 
$V=V_{\rm sad}(1+\etas y^2/2 + \sum_{n >1} c_n y^{2n})$, with the
coefficients $|c_n| \lesssim |\etas| y_{\rm end}^{2(1-n)}$ for $1<n<p$ , and
so negligible during inflation. Since $y_* \equiv y(N_*) \ll y_{\rm end}$, the
evolution of $y$ is then dominated by the $y^2$ and $y^{2p}$ terms. The
approximation~\eref{nsapprox} then generalises to
\be
n_s = 1+ 2 \etas -\frac{2(2p-1) \etas}{1-\rme^{-2(p-1)N \etas}} \, .
\label{nsp}
\ee
Figure~\ref{F:p} shows the above function for various values of $p$.
We see that as $p$ is increased, the bound on the spectral index is relaxed:
$n_s \lesssim 0.950, 0.968, 0.975, 0.980$ for $p=2,3,4,5$ respectively.
Larger values of $p$ require more fine-tuning, and hence progressively more
exponential terms are needed in the superpotential, in order
to have enough parameters to tune.

In \cite{race2} it was argued that anthropic considerations may 
favour models with the largest possible spectral index. 
In principle, double exponential superpotentials~\eref{racetrack1} contain
enough parameters to simultaneously fine-tune both $\etas$ and $C$ to be
small, as well as match the COBE normalisation for the power spectrum 
($P = V/[150 \pi^2 \epsilon] \approx 4 \times 10^{-10}$ at $N=N_*$). Hence 
$n_s \sim 0.97$ may be possible for the racetrack models~(\ref{racetrack1},
\ref{racetrack2}, \ref{racetrackD}). However, it will come at the cost of
even more severe fine-tuning.  

To summarise, in this paper we have reduced a general racetrack model to a
single field inflation model. This allowed us to derive a simple analytic
expression for the spectral index \eref{nsapprox}, and show that it is
bounded from above. Barring exceptional fine-tuning, i.e.\ any tuning
beyond that needed to get $\eta$ small, we predict $n_s \lesssim 0.95$ for all
racetrack inflation models in agreement with the WMAP3 data.

\end{document}